\documentclass[sigconf]{acmart}

\usepackage{balance}

\def\BibTeX{{\rm B\kern-.05em{\sc i\kern-.025em b}\kern-.08emT\kern-.1667em\lower.7ex\hbox{E}\kern-.125emX}}
    
\copyrightyear{2019}
\acmYear{2019}
\setcopyright{acmcopyright}
\acmConference[SIGIR '19] {42nd International ACM SIGIR Conference on Research and Development in Information Retrieval}{July 21--25, 2019}{Paris, France}
\acmBooktitle{42nd Int'l ACM SIGIR Conference on Research \& Development in Information Retrieval (SIGIR'19), July 21--25, 2019, Paris, France}
\acmPrice{15.00}
\acmDOI{10.1145/3331184.3331371}
\acmISBN{978-1-4503-6172-9/19/07}

\acmSubmissionID{sp556}

\begin{document}

\fancyhead{}

\title{NRPA: Neural Recommendation with Personalized Attention}

\author{Hongtao Liu}
\affiliation{%
  \institution{College of Intelligence and Computing, Tianjin University}
}
\email{htliu@tju.edu.cn}

\author{Fangzhao Wu}
\affiliation{%
  \institution{Microsoft Research Asia}
}
\email{wufangzhao@gmail.com}

\author{Wenjun Wang}
\affiliation{%
  \institution{College of Intelligence and Computing, Tianjin University}
}
\email{wjwang@tju.edu.cn}

\author{Xianchen Wang}
\affiliation{%
  \institution{College of Intelligence and Computing, Tianjin University}
}
\email{wangxc@tju.edu.cn}

\author{Pengfei Jiao}
\authornote{*Corresponding Author: Pengfei Jiao, pjiao@tju.edu.cn }
\affiliation{%
  \institution{Center for Biosafety Research and Strategy, Tianjin University}
}
\email{pjiao@tju.edu.cn}

\author{Chuhan Wu}
\affiliation{%
  \institution{Electronic Engineering}
  \city{Tsinghua University}
}
\email{wuch15@mails.tsinghua.edu.cn}

\author{Xing Xie}
\affiliation{%
  \institution{Microsoft Research Asia}
}
\email{xing.xie@microsoft.com}

%

\renewcommand{\shortauthors}{Hongtao Liu, Fangzhao Wu et al.}

%
\begin{abstract}

Existing review-based recommendation methods usually use the same model to learn the representations of all users/items from reviews posted by users towards items.
However, different users have different preference and different items have different characteristics.
Thus, the same word or the similar reviews may have different informativeness for  different users and items.
In this paper we propose a neural recommendation approach with personalized attention to learn personalized representations of users and items from reviews.
We use a review encoder to learn representations of reviews from words, and a user/item encoder to learn representations of users or items from reviews.
We propose a personalized attention model, and apply it to both review and user/item encoders to select different important words and reviews for different users/items.
Experiments on five datasets validate our approach can effectively improve the performance of neural recommendation.

\end{abstract}

%
%


%

%

%
\maketitle

\section{Introduction}
Recommender Systems (RS) are an information filtering systems that can learn user's interests and hobbies based on their historical behavior records, and predict users preference or ratings for items, which are ubiquitous today at e-commerce platforms such as Amazon and Netflix.

A number of works have been proposed for recommendation systems.
Collaborative Filtering (CF)~\cite{linden2003amazon} techniques are one of the most popular recommender methods, which are extensively used in industry.
Many of CF techniques are based on matrix factorization (MF) that decomposes the user-item rating matrix into two matrices corresponding to latent features of users and items~\cite{rendle2010factorization}.
However, these methods represent users and items only based on numeric ratings while the ratings suffer from the natural sparsity. 
Using text reviews to model user preference and item features is one approach to alleviate the above issues~\cite{kim2016convolutional,chen2018neural,lu2018coevolutionary,wang2019neural,wu2019hierarchical}. 
For example, 
ConvMF~\cite{kim2016convolutional} integrates convolutional neural network into probabilistic matrix factorization to exploit both ratings and item description documents.
TARMF~\cite{lu2018coevolutionary} utilizes attention-based recurrent neural networks to extract topical information from reviews and integrates textual features into probabilistic matrix factorization to enhance the performance of recommendation.

Despite their significant improvement of performance in recommendation, most existing methods learn  the representations from reviews  for all users or items using  the same model and ignore the deep personalized feature of users and items.
As a concrete example, suppose that User A cares more about the price of items than the quality and User B cares more quality than price, both of them write a similar review such as ``this  camera with a high price is easy to use.'' and then User A would give the camera a unsatisfied rating since the price is high while User B would vote a satisfied rating.
Thus, the same reviews are of different informativeness in terms of different users or items and it is necessary to be more personalized when learning representations from reviews for users or items.
As a result, we should exploit the individuality of users and items for neural recommendation. 

To this end, we propose a Neural Recommendation with hierarchical Personalized Attention (NRPA) model to learn  personalized representation for users and items.
Specifically, our NRPA contains two components, i.e., a review encoder to learn representations of reviews, a user/item encoder to learn representations of user/item from their reviews.
In review encoder, we utilize Convolutional Neural Network (CNN) to extract semantic features of reviews from words, and then  use personalized word-level attention to select more important words in a review for each user/item.
In user/item encoder we apply personalized review-level attention to learn the user/item representation via aggregating all the reviews representations according to their weights.
Moreover, the word- and review-level attention vectors of a user/item are generated by two multi-layer neural networks with the user/item ID embedding as input.
The two attention vectors can be seen as a indicator for each user and item under hierarchical views (i.e., word and review level).
At last,  we combine the representations of a user and a target item and feed them into a Factorization Machine~\cite{rendle2010factorization} layer to predict the rating that the user would vote the item.
We conduct extensive experiments on five benchmark datasets for recommendation with reviews.
The  results validate the  effectiveness of our personalized attention.

\section{Proposed Method}

\begin{figure}
\centering
\includegraphics[width=0.47\textwidth]{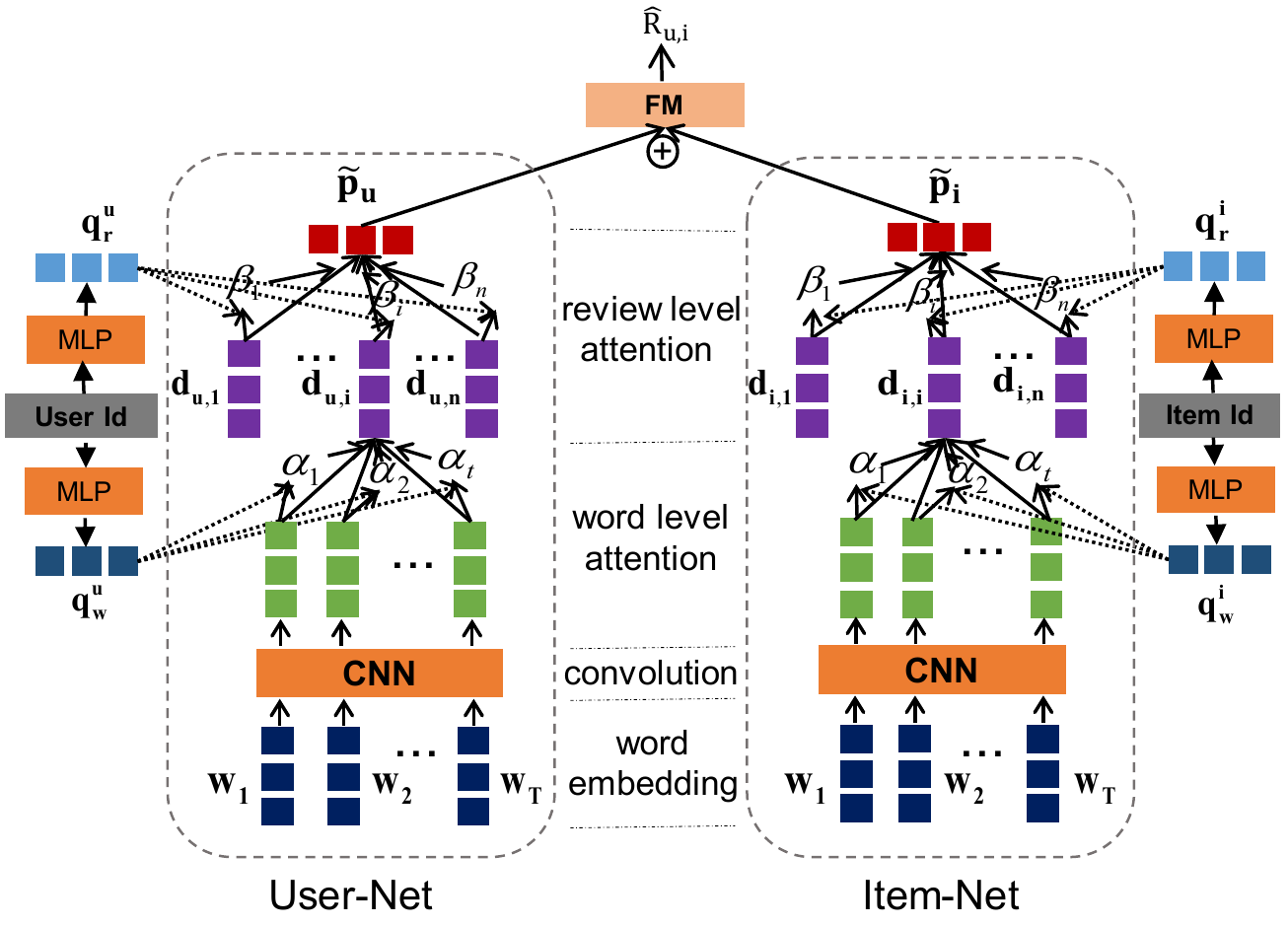}
\caption{\label{fig:overview} The framework of our \textit{NRPA} approach.}
\end{figure}

In this section, we introduce our NRPA approach in detail. 
Our approach contains three major components, i.e., a User-Net to learn user representations, an Item-Net to learn item representations, and a rating prediction module to predict the rating scores based on user and item representations.
Both User-Net and Item-Net contain two modules, i.e., a review encoder to learn representations of reviews from words and a user/item encoder to learn representations of users and items from reviews.
The overview of our NRPA approach is shown in Figure~\ref{fig:overview}. 

\subsection{Review Encoder}

We utilize word embedding to map each word into low-dimensional vectors and use Convolutional Neural Network (CNN) to extract the semantic features of text reviews. 
Then, we introduce a personalized attention into our model to highlight important words. 
Some notations used in the following section are defined as follows: 
$\mathit{U}$ and $\mathit{I}$ indicates the user set and item set, respectively.
$\mathbf{R} \in \mathcal{R}^{|U| \times |I|}$ denotes the rating matrix and $D \in \mathcal{R}^{|U| \times |I|}$ means the text review collection. 
$d_{u,i} = \{w_1, \cdots, w_T\}$ denotes the review written by user $u$ for item $i$.

\paragraph{Word Embedding and Convolution}
Given a review $d_{u,i}$, we first embed each word $w_k$ in $d_{u,i}$ to a $d_w$ dimensional vector $\mathbf{w_k}$ via word embedding.
Then, we transform the review $d_{u,i}$ into a matrix $\mathbf{M_{u,i}} = \left[ \mathbf{w_1, w_2, \cdots, w_T} \right]$. 
Afterwards, we perform convolution operator to calculate the feature matrix of the review $\mathbf{C} \in \mathcal{R}^{K \times T}$:
\begin{equation}
    \mathbf{C_j} = \sigma(\mathbf{W_j * M_{u,i} + b_j}) \ , 1 \leq j \leq K \ ,
\end{equation}
where $*$ is the convolution operator, $K$ is the number of filters and $\mathbf{W_j}$ is the weight matrix of the $j$-th filter.
Each column in $\mathbf{C}$ (denoted as $\mathbf{z_k} \in \mathcal{R}^{K}$) represents the semantic feature of the $k$-th word in the review.

\paragraph{Personalized Attention Vector over Word Level}
We first explore how to generate the attention vector $\mathbf{q^{u}_{w}}$ for the user $u$ which can embody the personalization.
Since each user or item has the unique id feature, we first represent all users and items into low-dimensional vectors via an embedding layer based on their IDs.
As shown in Figure~\ref{fig:overview}, given the id embedding of the user $u$, we utilize Multilayer Perceptron (MLP) to generate the personalized attention vector for user $u$, denoted as:
\begin{equation}
    \mathbf{q^{u}_w} = \mathtt{ReLU}(\mathbf{W_1u_{id}} + \mathbf{b_1}) \ ,
\end{equation}
where $\mathbf{W_1}$ is the weight matrix of MLP, $\mathbf{b_1}$ is the bias term and $\mathbf{u_{id}}$ is the ID embedding of user $u$.

\paragraph{User-Specific Attention over Word Level}
As mentioned above, not all words of a review are equally important for the representation of the review meaning. 
To highlight the important words, we employ the attention pooling mechanism in word level, denoted as:
\begin{equation}
    g_k = \mathbf{q^{u}_{w}Az_k} \ ,
\end{equation}
\begin{equation}
    \alpha_k = \frac{\exp(g_k)}{\sum_{j=1}^T \exp(g_j)}, \ \alpha_k \in (0, 1) \ ,
\end{equation}
where $\mathbf{A}$ is the harmony matrix in attention,
$\mathbf{q^{u}_{w}}$ is the attention vector specifically for the user $u$ obtained above, and $\mathbf{z_k}$ is the  representations of the $k$-th word above. 
$\alpha_i$ is attention weight of the $i$-th word in the review.
Similarly, the item $i$ has an unique attention vector denoted as $\mathbf{q^{i}_{w}}$ in Item-Net. 
Afterwards, we obtain the representation of the $i$-th review of user $u$ via aggregating feature vectors of all words:
\begin{equation}
    \mathbf{d_{u,i}}=\sum_{j=1}^{T}\alpha_j \mathbf{z_j} \ .
\end{equation}

\subsection{User and Item Encoder}
After we have obtained all the review representations of users and items above, we will explore how to aggregate them together to represent users or items.
As stated above, different reviews are of different importance for representation of user.
Besides, the information of a review varies from different users. 
Hence, we introduce a user-specific attention mechanism to focus on the useful reviews for each user.

\paragraph{Personalized Attention Vectors over Review Level} \quad 
Based on the user id embedding $\mathbf{u_{id}}$, we utilize another MLP layer to generate the personalized review-level attention vector for user $u$:
\begin{equation}
    \mathbf{q^{u}_r} = \mathtt{ReLU}(\mathbf{W_2u_{id}} + \mathbf{b_2}) \ ,
\end{equation}
where $\mathtt{ReLU}$ is the activation function, $\mathbf{W_2}$ is the weight matrix, $\mathbf{b_2}$ is the bias term.

\paragraph{User-Specific Attention over Review Level} \quad Given the review set $d_{u} = \{ {d_{u,1}, d_{u,2}, \cdots, d_{u,N}} \}$, we apply attention to highlight those informative reviews and de-emphasize those meaningless. 
To be specific, we compute the weight $\beta_j$ of the $j$-th review of the $i$-th user as follows:
\begin{equation}
   e_j = \mathbf{q^{u}_rA_2d_{u,j}} \ ,
\end{equation}
\begin{equation}
    \beta_j = \frac{\exp(e_j)}{\sum_{k=1}^{N}\exp(e_k)}, \ \beta_j \in (0,1) \ ,
\end{equation}
where $\mathbf{A_2}$ is the  matrix in attention; $\mathbf{q^{u}_r}$ is the query vector for the  user $u$ and each users have a unique attention vector to find  informative reviews. 
Afterwards, we obtain the text feature $\mathbf{\tilde{p}_u}$ of user $u$ via aggregating all the reviews according to their weights:
\begin{equation}
    \mathbf{\tilde{p}_u} = \sum_{j=1}^{N} \beta_j \mathbf{d_{u,j}} \ .
\end{equation}

Similarly, we can get the feature of item $i$, denoted as $\mathbf{\tilde{p}_i}$.

\subsection{Rating Prediction}

In this section we predict the ratings based on $\mathbf{\tilde{p}_u}$ and $\mathbf{\tilde{p}_i}$.
First, we concatenate $\mathbf{\tilde{p}_u}$ and $\mathbf{\tilde{p}_i}$ and feed into a Factorization Machine (FM)~\cite{rendle2010factorization} to predict rating:
\begin{equation}
    \mathbf{\hat{o} = \tilde{p}_u \oplus \tilde{p}_i} \ ,
\end{equation}

\begin{equation}
        \hat{R}_{u,i} = \hat{w}_0 + \sum^{|\mathbf{\hat{o}}|}_{i=1}\hat{w}_i \hat{o}_i + \sum^{|\mathbf{\hat{o}}|}_{i=1}\sum^{|\mathbf{\hat{o}}|}_{j=i+1}\left\langle \mathbf{\hat{v}_i,\hat{v}_j}\right\rangle \hat{o}_i\hat{o}_j \ ,
\end{equation}
where $\oplus$ is the concatenation operation; $\hat{w}_0$ and $\hat{w}_i$ are both parameters in FM.


\section{Experiments}

\subsection{Datasets and Experimental Settings}

Our experiments are conducted on five benchmark datasets.
The first two datasets Yelp 2013 (denoted as Yelp13) and Yelp 2014 (denoted as Yelp14) are selected from Yelp Dataset Challenge\footnote{https://www.yelp.com/dataset/challenge}.
The other three datasets Electronics, Video Games and Gourmet Foods are selected from Amazon dataset\footnote{http://jmcauley.ucsd.edu/data/amazon/}, and we denote them as Elec., Games and Foods respectively.
Note that all datasets contain reviews with ratings (from 1 to 5).
The details of the datasets are shown in Table~\ref{tab:dataset}.
We randomly split each dataset into training set, validation set and test set with 80\%, 10\% and 10\% respectively. 

In our experiments, we use validation dataset to tune the hyperparameters in our model.
The word embedding vectors are 300-dimensional.
The dimension of ID embedding is set to 32.
The number of filters in CNN and the dimension of attention vectors is 80.
The window size of CNN is set to 3.
Following previous works~\cite{zheng2017joint,chen2018neural}, we utilize  Mean Squared Error (MSE) as the evaluation metric.

\begin{table}
\centering
\caption{Statistics of the five datasets in our experiments.}
\label{tab:dataset}
\resizebox{0.34\textwidth}{!}{
\begin{tabular}{|c|c|c|c|c|}
\hline
Dataset &  \#users & \#items & \#ratings & density\\
\hline
Yelp13 & 1,631 & 1,633 & 78,966 & 2.965 \\
Yelp14 & 4,818 & 4,194 & 231,163 & 1.144 \\
Elec. & 192,403 & 63,001 & 1,689,188 & 0.014 \\
Games & 24,303 & 10,672 & 231,780 & 0.089 \\
Foods & 14,681 & 8,713 & 151,254 & 0.118 \\
\hline
\end{tabular}
}
\end{table}

\subsection{Performance Evaluation}

We evaluate our method NRPA with the following  baseline methods:
\begin{itemize}
\item \textbf{PMF}~\cite{mnih2008probabilistic} models the latent factors for users and items by introducing Gaussian distribution.
\item \textbf{CTR}~\cite{wang2011collaborative} learns interpretable latent structure from user-generated content to integrate probabilistic modeling into collaborative filtering.
\item \textbf{ConvMF+}~\cite{kim2016convolutional} incorporates  convolutional neural network into Matrix Factorization  to learn item features from item review documents.
\item \textbf{DeepCoNN}~\cite{zheng2017joint}  models users and items via combining all their associated reviews by convolutional neural network.
\item \textbf{NARRE}~\cite{chen2018neural} is a newly proposed method that introduces neural attention mechanism to build the recommendation model and select highly-useful reviews simultaneously.
\item \textbf{TARMF}~\cite{lu2018coevolutionary} is a recommendation model which utilizes attention-based recurrent neural networks to extract topical information from review documents. 
\end{itemize}

\begin{table}
\caption{\textbf{Comparisons between NRPA and baselines. }}\label{tab:results}
\centering
\resizebox{0.4\textwidth}{!}{
\begin{tabular}{c|c|c|c|c|c}
\hline
  & Yelp13 & Yelp14 & Elec. & Games & Foods\\
\hline
PMF & 0.985 & 1.053 & 1.411 & 1.297 & 1.251\\
CTR & 0.975 & 1.013 & 1.284 & 1.147 & 1.139\\
\hline
ConvMF+ & 0.917 & 0.954 & 1.241 & 1.092 & 1.084\\
DeepCoNN & 0.880 & 0.910 & 1.232 & 1.130 & 0.985  \\
NARRE & 0.879 & 0.906 & 1.215 & 1.112 & 0.986 \\
TARMF &  0.875 & 0.909 & 1.147 & 1.043 & 1.019\\
\hline
NRPA & \bf 0.872 & \bf 0.897 & \bf 1.047 & \bf1.014 & \bf 0.953\\
\hline
\end{tabular}
}
\end{table}

The MSE results of all methods are shown in Table~\ref{tab:results}.
Our model NRPA outperforms all the baseline methods among all the five datasets which indicates the robust effectiveness of our personalized attention in modeling users and items.
Besides, we can observe that (1) The methods with reviews perform better than those methods with only ratings (i.e., PMF and CTR).
The reason may be that the reviews with the rich semantic textual information are powerful in capturing the feature of users and items.
(2) Though both NARRE and TARMF utilize the attention mechanism to focus on more important information, our method NRPA achieves a better performance than them. 
We conclude that our model NRPA with hierarchical attention can exploit the deep personalized features (i.e., word level and review level) of users and items, which can represent users and items more precise.
This result is consistent with our intuitive motivation that users and items should be characterized by individuation in recommendation.

\subsection{Effectiveness of Personalized Attention}
\begin{figure}
\centering
\includegraphics[width=0.32\textwidth]{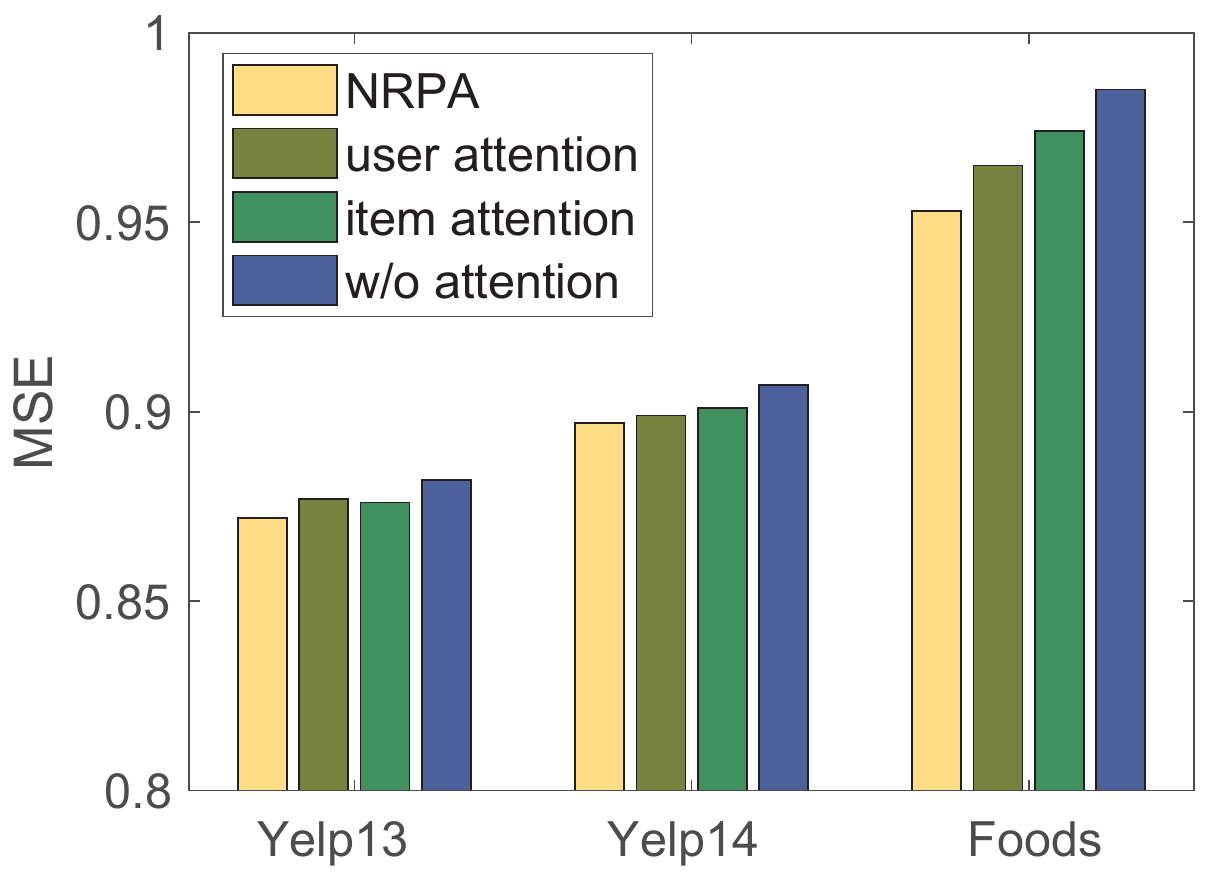}
\caption{\label{fig:mse_att} Effectiveness of user and item attentions.}
\end{figure}

In this section we further explore the effectiveness of our personalized attention module.
First we evaluate the effect of user attention and item attention respectively.
From the results in Figure~\ref{fig:mse_att} (experiments on three datasets for space limitation), we can find that comparing with the variant without attention (i.e., average weight for all reviews), both variants with user attention and item attention can improve the performance of rating prediction in recommendation.
This is because different users or items always have their unique preference or features.
And our personalized attention can effectively capture the personality of users and items, which is beneficial for learning a precise representation of users and items.

\begin{figure}
\centering
\includegraphics[width=0.32\textwidth]{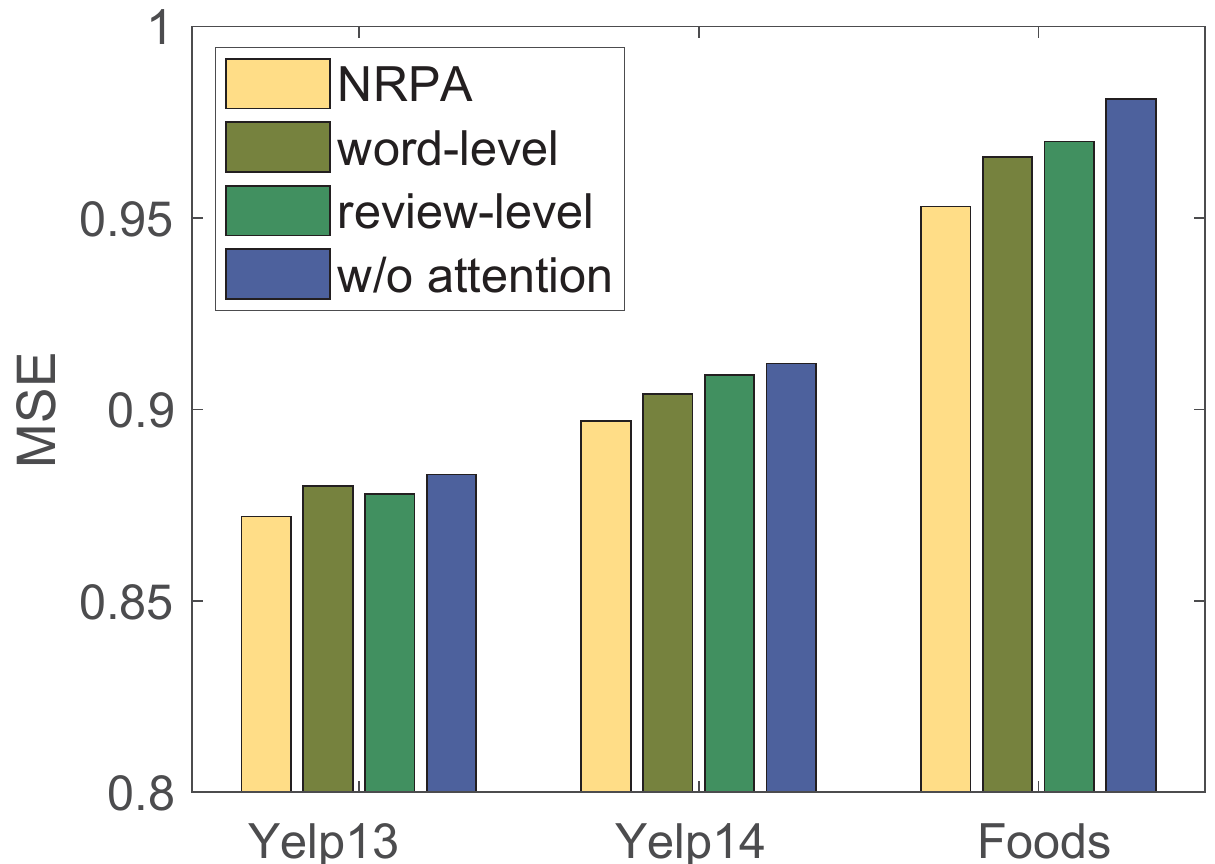}
\caption{\label{fig:mse_wr} Effectiveness of word- and review-level attentions.}
\end{figure}

Besides, we explore the effectiveness of word-level attention and review-level attention.
As shown in Figure~\ref{fig:mse_wr}, we can observe that the variants with only word-level attention and only review-level attention can  both perform better than the model without any attention.
This is because word level attention can recognize those important words for each user or item;
review level attention can help to focus on the more informative reviews during modeling user preference and item features.

\subsection{Parameter Analysis}
\begin{figure}
\centering
\includegraphics[width=0.34\textwidth]{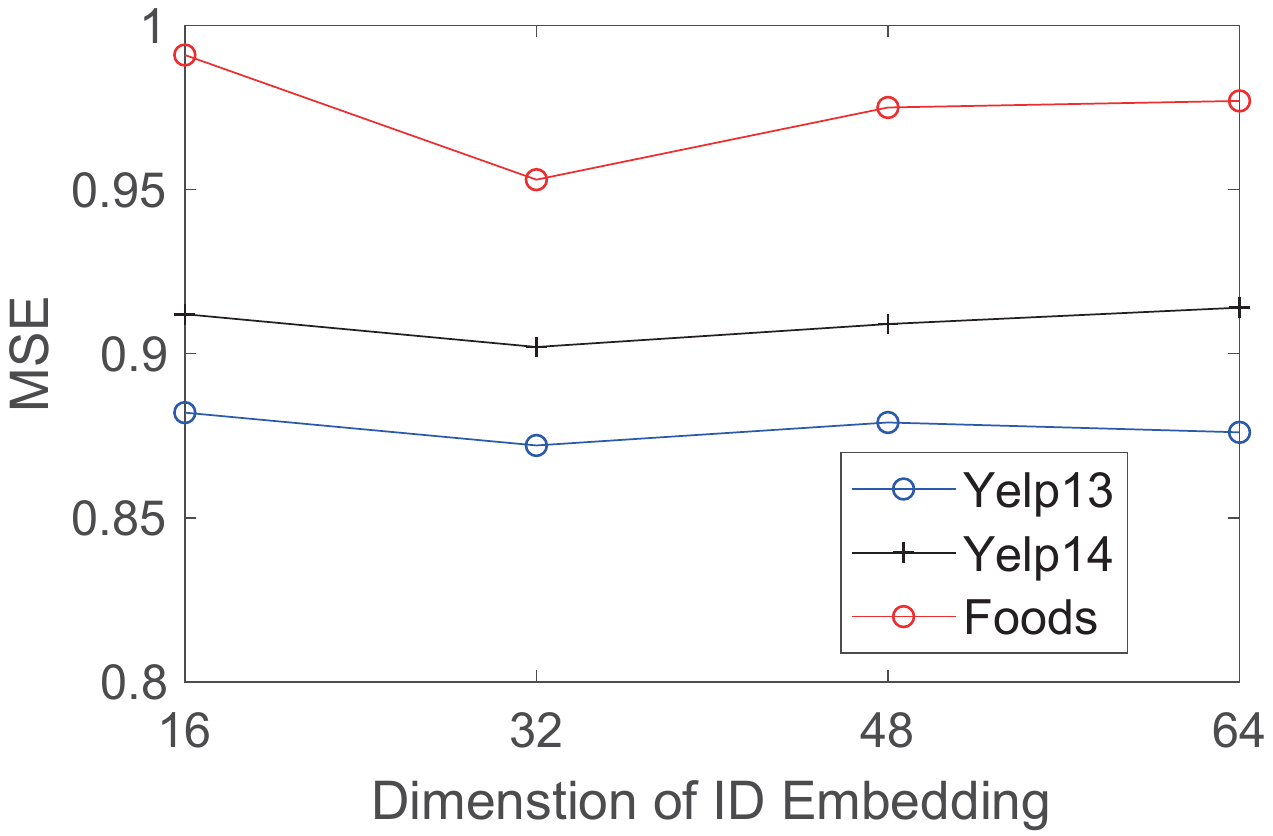}
\caption{\label{fig:para} The influence of ID embedding dimension.}
\end{figure}

Since our personalized attention vectors are generated by the user and item id embedding, this section explores the effect of varying dimension of the id embedding.
From the result in Figure~\ref{fig:para}, we can see that as the dimension increases,  the  MSE first decrease, then reaches the best, and decreases afterwards. 
When dimension is too smaller,  the attention vectors may  not learn the diversity of users and items enough. 
However if the dimension becomes too large, the model may suffer from overfitting.
The optimal value of the ID dimension is 32 regardless of different datasets.

\section{Conclusion}
In this paper, we propose a neural recommendation approach with personalized attentions to learn personalized user and item representations from reviews.
The core of our approach is a personalized attention model whose query vectors are learned from the embeddings of user and item IDs.
We apply this attention model to both review encoder and user/item encoder to select different important words and reviews for different users and items.
In this way the different preference of users and different characteristics of items can be better captured.
The experiments on five benchmark datasets show that our approach can effectively improve the performance of neural recommendation.

%

\begin{acks}
This work was supported by the National Key R\&D Program of China (2018YFC0809800, 2018YFC0831000), the National Natural Science Foundation of China (91746205, 91746107, 51438009).
\end{acks}

\bibliographystyle{ACM-Reference-Format}
\bibliography{sigir2019.bib}


\begin{thebibliography}{10}


\ifx \showCODEN    \undefined \def \showCODEN     #1{\unskip}     \fi
\ifx \showDOI      \undefined \def \showDOI       #1{#1}\fi
\ifx \showISBNx    \undefined \def \showISBNx     #1{\unskip}     \fi
\ifx \showISBNxiii \undefined \def \showISBNxiii  #1{\unskip}     \fi
\ifx \showISSN     \undefined \def \showISSN      #1{\unskip}     \fi
\ifx \showLCCN     \undefined \def \showLCCN      #1{\unskip}     \fi
\ifx \shownote     \undefined \def \shownote      #1{#1}          \fi
\ifx \showarticletitle \undefined \def \showarticletitle #1{#1}   \fi
\ifx \showURL      \undefined \def \showURL       {\relax}        \fi
\providecommand\bibfield[2]{#2}
\providecommand\bibinfo[2]{#2}
\providecommand\natexlab[1]{#1}
\providecommand\showeprint[2][]{arXiv:#2}

\bibitem[\protect\citeauthoryear{Chen, Zhang, Liu, and Ma}{Chen
  et~al\mbox{.}}{2018}]%
        {chen2018neural}
\bibfield{author}{\bibinfo{person}{Chong Chen}, \bibinfo{person}{Min Zhang},
  \bibinfo{person}{Yiqun Liu}, {and} \bibinfo{person}{Shaoping Ma}.}
  \bibinfo{year}{2018}\natexlab{}.
\newblock \showarticletitle{Neural attentional rating regression with
  review-level explanations}. In \bibinfo{booktitle}{\emph{WWW}}.
  \bibinfo{pages}{1583--1592}.
\newblock


\bibitem[\protect\citeauthoryear{Kim, Park, Oh, Lee, and Yu}{Kim
  et~al\mbox{.}}{2016}]%
        {kim2016convolutional}
\bibfield{author}{\bibinfo{person}{Donghyun Kim}, \bibinfo{person}{Chanyoung
  Park}, \bibinfo{person}{Jinoh Oh}, \bibinfo{person}{Sungyoung Lee}, {and}
  \bibinfo{person}{Hwanjo Yu}.} \bibinfo{year}{2016}\natexlab{}.
\newblock \showarticletitle{Convolutional matrix factorization for document
  context-aware recommendation}. In \bibinfo{booktitle}{\emph{RecSys}}.
  \bibinfo{pages}{233--240}.
\newblock


\bibitem[\protect\citeauthoryear{Linden, Smith, and York}{Linden
  et~al\mbox{.}}{2003}]%
        {linden2003amazon}
\bibfield{author}{\bibinfo{person}{Greg Linden}, \bibinfo{person}{Brent Smith},
  {and} \bibinfo{person}{Jeremy York}.} \bibinfo{year}{2003}\natexlab{}.
\newblock \showarticletitle{Amazon.com recommendations: Item-to-item
  collaborative filtering}.
\newblock \bibinfo{journal}{\emph{Internet Computing}} (\bibinfo{year}{2003}),
  \bibinfo{pages}{76--80}.
\newblock


\bibitem[\protect\citeauthoryear{Lu, Dong, and Smyth}{Lu et~al\mbox{.}}{2018}]%
        {lu2018coevolutionary}
\bibfield{author}{\bibinfo{person}{Yichao Lu}, \bibinfo{person}{Ruihai Dong},
  {and} \bibinfo{person}{Barry Smyth}.} \bibinfo{year}{2018}\natexlab{}.
\newblock \showarticletitle{Coevolutionary Recommendation Model: Mutual
  Learning between Ratings and Reviews}. In \bibinfo{booktitle}{\emph{WWW}}.
  \bibinfo{pages}{773--782}.
\newblock


\bibitem[\protect\citeauthoryear{Mnih and Salakhutdinov}{Mnih and
  Salakhutdinov}{2008}]%
        {mnih2008probabilistic}
\bibfield{author}{\bibinfo{person}{Andriy Mnih} {and} \bibinfo{person}{Ruslan~R
  Salakhutdinov}.} \bibinfo{year}{2008}\natexlab{}.
\newblock \showarticletitle{Probabilistic matrix factorization}. In
  \bibinfo{booktitle}{\emph{NIPS}}. \bibinfo{pages}{1257--1264}.
\newblock


\bibitem[\protect\citeauthoryear{Rendle}{Rendle}{2010}]%
        {rendle2010factorization}
\bibfield{author}{\bibinfo{person}{Steffen Rendle}.}
  \bibinfo{year}{2010}\natexlab{}.
\newblock \showarticletitle{Factorization machines}. In
  \bibinfo{booktitle}{\emph{ICDM}}. \bibinfo{pages}{995--1000}.
\newblock


\bibitem[\protect\citeauthoryear{Wang and Blei}{Wang and Blei}{2011}]%
        {wang2011collaborative}
\bibfield{author}{\bibinfo{person}{Chong Wang} {and} \bibinfo{person}{David~M
  Blei}.} \bibinfo{year}{2011}\natexlab{}.
\newblock \showarticletitle{Collaborative topic modeling for recommending
  scientific articles}. In \bibinfo{booktitle}{\emph{KDD}}.
  \bibinfo{pages}{448--456}.
\newblock


\bibitem[\protect\citeauthoryear{Wang, Liu, Wang, Wu, Xu, Wang, and Xie}{Wang
  et~al\mbox{.}}{2019}]%
        {wang2019neural}
\bibfield{author}{\bibinfo{person}{Xianchen Wang}, \bibinfo{person}{Hongtao
  Liu}, \bibinfo{person}{Peiyi Wang}, \bibinfo{person}{Fangzhao Wu},
  \bibinfo{person}{Hongyan Xu}, \bibinfo{person}{Wenjun Wang}, {and}
  \bibinfo{person}{Xing Xie}.} \bibinfo{year}{2019}\natexlab{}.
\newblock \showarticletitle{Neural Review Rating Prediction with Hierarchical
  Attentions and Latent Factors}. In \bibinfo{booktitle}{\emph{DASFAA}}.
  \bibinfo{pages}{363--367}.
\newblock


\bibitem[\protect\citeauthoryear{Wu, Wu, Liu, and Huang}{Wu
  et~al\mbox{.}}{2019}]%
        {wu2019hierarchical}
\bibfield{author}{\bibinfo{person}{Chuhan Wu}, \bibinfo{person}{Fangzhao Wu},
  \bibinfo{person}{Junxin Liu}, {and} \bibinfo{person}{Yongfeng Huang}.}
  \bibinfo{year}{2019}\natexlab{}.
\newblock \showarticletitle{Hierarchical User and Item Representation with
  Three-Tier Attention for Recommendation}. In
  \bibinfo{booktitle}{\emph{NAACL}}.
\newblock


\bibitem[\protect\citeauthoryear{Zheng, Noroozi, and Yu}{Zheng
  et~al\mbox{.}}{2017}]%
        {zheng2017joint}
\bibfield{author}{\bibinfo{person}{Lei Zheng}, \bibinfo{person}{Vahid Noroozi},
  {and} \bibinfo{person}{Philip~S Yu}.} \bibinfo{year}{2017}\natexlab{}.
\newblock \showarticletitle{Joint deep modeling of users and items using
  reviews for recommendation}. In \bibinfo{booktitle}{\emph{WSDM}}.
  \bibinfo{pages}{425--434}.
\newblock


\end{thebibliography}

\end{document}